\documentclass{Interspeech}
\usepackage{multirow}
\usepackage{makecell} 
\usepackage{bbding,pifont}
\usepackage{adjustbox}
\usepackage{amssymb}
\interspeechcameraready 
\title{SpeechRefiner: Towards Perceptual Quality Refinement for Front-End Algorithms
\thanks{*:  Corresponding author.}}
\author[affiliation={1}]{Sirui}{Li}
\author[affiliation={2,1,3,*}]{Shuai}{Wang}
\author[affiliation={1}]{Zhijun}{Liu}
\author[affiliation={4}]{Zhongjie}{Jiang}
\author[affiliation={4}]{Yannan}{Wang}
\author[affiliation={1,3}]{Haizhou}{Li}
\affiliation{School of Data Science}{The Chinese University of Hong Kong, Shenzhen}{China}
\affiliation{School of Intelligence Science and Technology}{Nanjing University, Suzhou}{China}
\affiliation{}{Shenzhen Research Institute of Big Data, Shenzhen}{China}
\affiliation{Tencent Ethereal Audio Lab}{Tencent, Shenzhen}{China}
\email{223040027@link.cuhk.edu.cn, shuaiwang@nju.edu.cn}

\usepackage{comment}
\begin{document}
\ninept
\maketitle
\begin{abstract}

Speech pre-processing techniques such as denoising, de-reverberation, and separation, are commonly employed as front-ends for various downstream speech processing tasks. However, these methods can sometimes be inadequate, resulting in residual noise or the introduction of new artifacts. Such deficiencies are typically not captured by metrics like SI-SNR but are noticeable to human listeners. To address this, we introduce SpeechRefiner, a post-processing tool that utilizes Conditional Flow Matching (CFM) to improve the perceptual quality of speech. In this study, we benchmark SpeechRefiner against recent task-specific refinement methods and evaluate its performance within our internal processing pipeline, which integrates multiple front-end algorithms. Experiments show that SpeechRefiner exhibits strong generalization across diverse impairment sources, significantly enhancing speech perceptual quality. Audio demos can be found at \url{https://speechrefiner.github.io/SpeechRefiner/}.

\end{abstract}
\begin{figure*}[ht!]
    \centering
    \includegraphics[width=0.99\textwidth]{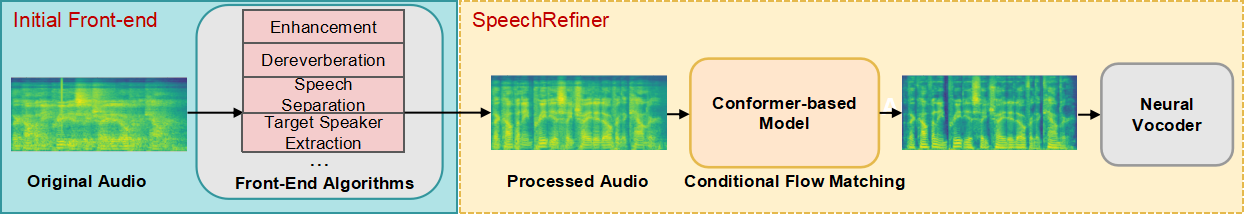}
    \caption{Overview of the SpeechRefiner Augmented Speech Restoration System. SpeechRefiner acts as a post-processing module to address the distortions and artifacts introduced by front-end processing algorithms. 
}\label{fig:system_overview}
\end{figure*}

\begin{figure}[!htb]
\centering
\includegraphics[width=\linewidth, height=2\textheight, keepaspectratio]{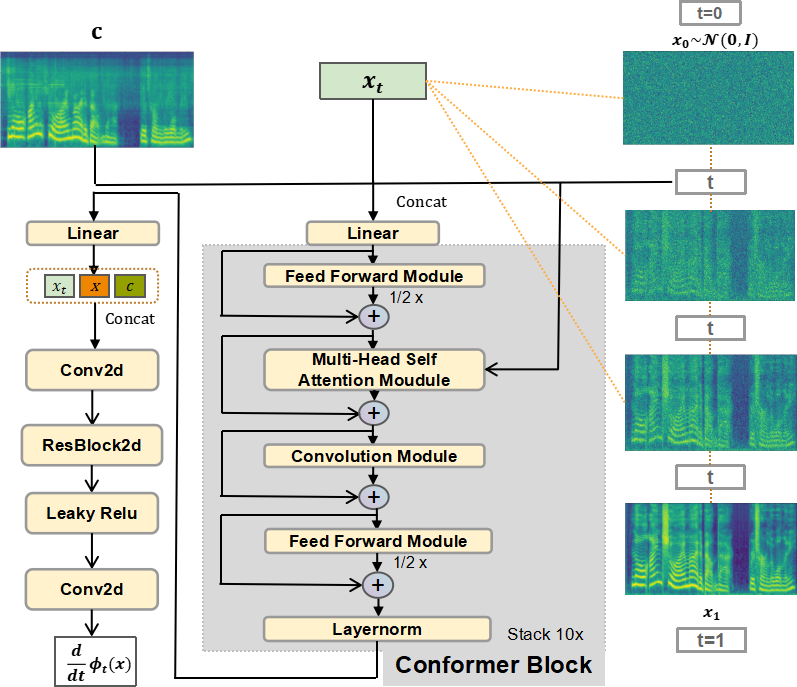}
\caption{The architecture of the conformer-based model.}
\label{fig:archi}
\end{figure}
\section{Introduction}
\label{sec:related}
Speech pre-processing front-ends are critical for intelligent systems functioning in real-world acoustic environments, where target human voices are often corrupted by reverberation, noise, and overlapping sound sources. Typical tasks involve enhancement techniques to reduce non-speech interference, such as de-noising~\cite{azarang2020review} and de-reverberation~\cite{naylor2010speech}, and separation techniques for multi-speaker scenarios, including blind speech separation~\cite{wang2018supervised,luo2019conv} and target speaker extraction~\cite{zmolikova2023neural}.

The evaluation of these  front-end algorithms predominantly relies on objective metrics like SI-SNR (Scale-Invariant Signal-to-Noise Ratio)~\cite{le2019sdr}, which compares processed signals with clean references. While SI-SNR optimization has driven performance improvements~\cite{luo2018tasnet}, this metric doesn't perfectly align with human perception.  A significant challenge is that front-end pre-processing can introduce artifacts, distortions, or spectral blurring, negatively impacting the auditory experience \cite{liu2024rad,liu2024rad2}---a critical issue for applications like smart conferencing and hearing aids where perceptual fidelity is essential.
This challenge also intersects with the era of large-scale data processing. Automated pipelines for cleaning in-the-wild audio data~\cite{yu2024autoprep,ma2024wenetspeech4tts} often discard substantial portions of raw datasets after quality filtering (e.g., using DNSMOS~\cite{reddy2021dnsmos} or SIGMOS~\cite{Ristea2024}). Integrating front-end enhancement into such pipelines could significantly improve retention of usable data, creating synergies between algorithmic advances and data-driven applications.

To address deficiencies in front-end processing, one approach is to employ speech restoration techniques to repair audio signals degraded by various impairments. Advanced methods such as VoiceFixer \cite{Liu2021VoiceFixer} and Miipher \cite{Koizumi2023Miipher} enhance traditional front-end capabilities by tackling complex degradation scenarios using sophisticated neural architectures. These restoration techniques aim to holistically improve the perceptual quality of the original audio, rather than specifically addressing the challenges introduced by the aforementioned front-end processing. 
In addition to these universal restoration methods, task-specific solutions have been developed to counteract the challenges introduced by front-end processing. These include additional refining modules tailored for speech enhancement \cite{lemercier2023storm, sawata2022diffiner}, dereverberation \cite{lemercier2023storm}, and separation \cite{hirano2023diffusion, lutati2023separate, wang2024noise}. While effective in their specific domains, these solutions are often closely integrated with particular front-end algorithms. For example, StoRM \cite{lemercier2023storm} and Fast-GeCo \cite{wang2024noise} necessitate joint training with front-end processors, and Diffiner \cite{sawata2022diffiner} relies on noise estimates from speech enhancement front-ends as supplementary inputs. This reliance on co-trained components and front-end-specific features increases computational complexity and reduces adaptability, thereby limiting their generalizability and utility across varied scenarios.

To overcome these limitations, we introduce SpeechRefiner, a novel audio restoration model utilizing Conditional Flow Matching (CFM) \cite{Lipman2023}. Unlike aforementioned approaches that often necessitate front-end-specific inputs or co-training, SpeechRefiner serves as a versatile post-processing solution that enhances the perceptual quality of audio signals from various front-end algorithms. 
We validated SpeechRefiner on speech processed by various front-ends, including enhancement, dereverberation, separation, and TSE. We first compared it against task-specific refinement models, then tested it on an internal front-end pipeline, which integrates multiple front-end algorithms. The results demonstrate that SpeechRefiner significantly improves perceptual quality and exhibits
strong generalization performance even on unseen front-end algorithms.

\section{SpeechRefiner}
In this work, we propose SpeechRefiner, a standalone refinement module designed to address degradation caused by various speech processing front-ends.

\subsection{Model overview}
SpeechRefiner consists of two main components, as shown in Figure~\ref{fig:system_overview}. First, a Conformer-based model predicts mel-spectrograms from the distorted speech output of the front-end processing algorithm. Then, a neural vocoder synthesizes the mel-spectrograms to restore the speech signal.

\subsection{Conformer-based Model}

\subsubsection{Optimal-Transport Conditional Flow Matching}
\label{ssec:flow_matching_intro}
Modeling data distributions in high-dimensional spaces is non-trival for neural networks, especially when the target true distribution \(q(x)\) is unknown. One feasible solution is transforming a simple prior distribution \(p_0\), such as a Gaussian, into a more complex target distribution \(p_1\), which approximates \(q(x)\). Flow matching ~\cite{Lipman2023} creates a continuous mapping, \(\phi_t: \mathbb{R}^d \rightarrow \mathbb{R}^d\), for \(t \in [0, 1]\), which progressively shifts the prior distribution to the target by learning a  time-dependent vector field $u_t: \mathbb{R}^d \to \mathbb{R}^d$. This transformation follows the dynamics defined by the ordinary differential equation (ODE):
\begin{align}
\frac{d}{dt} \phi_t(x) &= v_t\left(\phi_t(x)\right);\phi_0(x)= x.
\end{align}
However, directly calculating the vector field \(u_t\) and marginal density \(p_t\) is often computationally impractical, rendering flow matching difficult to apply. Conditional Flow Matching (CFM) \cite{Lipman2023} offers a solution by replacing the original vector field \(u_t\) with a conditional vector field \(u_t(x|x_1)\), making the problem more manageable. The loss function for CFM is expressed as:
\begin{align}
L_{CFM}(\theta) = \mathbb{E}_{t, q(x_1), p_t(x|x_1)}\Vert v_t(x; \theta) - u_t(x|x_1)\Vert^2.
\end{align}
Here, \(p_t(x|x_1)\) represents the conditional distribution at time \(t\), while \(v_t(x; \theta)\) is a neural network that approximates \(u_t(x|x_1)\). By leveraging conditional probabilities and fields, CFM avoids the difficulties of working with full marginal distributions. This approach simplifies computations and ensures consistent gradient estimates, facilitating efficient training.

Building on the CFM framework, our model using Optimal-Transport Conditional Flow Matching (OT-CFM) \cite{Lipman2023}, a variant of CFM that offers particularly straightforward gradients. The loss function for our model is defined as:
\begin{align}
L(\theta) = \mathbb{E}_{t,q(x_1),p_0(x_0)} \Vert v_t(\phi_t(x) | c; \theta) - u_t(x_0 | x_1) \Vert^2.
\end{align}
In this case, \(\phi_t(x) = (1 - (1 - \sigma_{\text{min}})t)x_0 + t x_1\) describes the flow from \(x_0\) to \(x_1\), each \(x_1\) paired with a sample \(x_0 \sim N(0, I)\). Additionally, the learning target's gradient field \(u_t(x_0|x_1) = x_1 - (1 - \sigma_{\min})x_0\). For our model, \(x_1\) refers to clean speech, while \(c\) denotes distorted speech. The parameter \(\sigma_{\text{min}}\) is a small-valued hyperparameter set to zero in our experiments. The vector field \(v_t(\phi_t(x) | c; \theta)\), which is the final output of our model, corresponds to the derivative \(\frac{d}{dt} \phi_t(x)\) mentioned in the ordinary differential equation. This equation indicates how the randomly sampled Gaussian noise \(x_0 \sim N(0, I)\) evolves at each time step \(t\) to match the target clean speech distribution. Essentially, this represents the slope of the flow at each time step.

\subsubsection{The Conformer Architecture}
The architecture of our conformer-based model is shown in Figure ~\ref{fig:archi}. The model takes \( x_t \) defined as the \(\phi_t(x)\) introduced above, along with additional conditioning features \( c \in \mathbb{R}^{N \times D_1 \times D_2} \) representing distorted speech and the current time step \( t \). Here, \( N \) denotes the batch size, \( D_1 \) represents the number of frames, and \( D_2 \) is the number of Mel frequency bins. These inputs are concatenated and passed to Conformer blocks.

Each Conformer block~\cite{Gulati2020} includes a feed-forward module, self-attention module, convolution module, and another feed-forward module. To capture long-range dependencies, we repeat this block 10 times. We replace the original relative position encoding with rotational position embedding (RoPE)\cite{Su2021}, which is computationally efficient and generalizes better to longer sequences\cite{Press2022}.

The Conformer output is transposed and concatenated with $x_t$ and \(c\), then processed by Conv2D and ResBlock2D layers. Each ResBlock2D contains two residual blocks with 2D convolutions and LeakyReLU activations. A final convolutional layer reduces the feature dimensions, producing the output \(\frac{d}{dt} \phi_t(x) \in \mathbb{R}^{N \times D_1 \times D_2}\).

\subsection{Neural Vocoder}
After obtaining the refined mel-spectrogram, we use a pretrained neural vocoder to reconstruct waveforms. Specifically, we adopt Vocos\footnote{\url{https://github.com/gemelo-ai/vocos}}\cite{Siuzdak2023}, which achieves state-of-the-art audio quality while significantly improving computational efficiency.

\section{Experimental Setups}
\subsection{Front-end processing systems}
We aim to test our system on speech processed by different front-end algorithms to evaluate its generalization performance.  For each task,
we randomly select 500 utterances from the corresponding test set to
evaluate the proposed SpeechRefiner. The related front-end algorithms and the corresponding training data are summarized in Table~\ref{tab:frontend}, covering tasks such as speech enhancement, speech dereverberation, and speech separation, all of which have existing restoration modules proposed by previous studies. Moreover, to enable a more comprehensive comparative analysis, we also include two target speaker extraction systems (Audio based and Audio-Visual based).

\begin{table}[h]
\centering
\caption{Front-end algorithms and their corresponding datasets.}
\label{tab:frontend}
\begin{adjustbox}{width=0.5\textwidth}
\begin{tabular}{l l l l}
\toprule
\multirow{2}{*}{\textbf{Task}} & \textbf{Front-end} & \textbf{Refine} & \textbf{Training} \\
                      & \textbf{Algorithm} & \textbf{Algorithm} & \textbf{Data} \\\midrule
Enhancement   & CDiffuSE~\cite{lu2022conditional}            & Diffiner+~\cite{sawata2022diffiner} & VB-DMD~\cite{valentini2016investigating}     \\ 
Dereverb      & SGMSE+~\cite{Richter2023}                    & StoRM~\cite{lemercier2023storm}    & WSJ0-REVERB~\cite{Richter2023} \\ 
Separation    & Sepformer~\cite{subakan2021attention}        & Fast-GeCo~\cite{wang2024noise}     & Libri2Mix~\cite{cosentino2020librimix} \\ \midrule
Audio-TSE     & Spex+~\cite{ge2020spex+}                     & -                                  & Libri2Mix~\cite{cosentino2020librimix} \\ 
AV-TSE        & MuSE~\cite{pan2021muse}                      & -                                  & Vox2-Mix~\cite{pan2021muse} \\ 
\bottomrule
\end{tabular}
\end{adjustbox}
\end{table}

Specifically, both CDiffuSE\footnote{\url{https://github.com/neillu23/CDiffuSE}} and SGMSE+\footnote{\url{https://github.com/sp-uhh/sgmse}} are trained using the official implementations on the respective datasets, while Sepformer utilizes the checkpoint provided in the Fast-GeCo repository\footnote{\url{https://github.com/WangHelin1997/Fast-GeCo}}.

\subsection{Baseline refinement systems}

For the baseline systems, we first adapt the well-known restoration toolkit VoiceFixer~\cite{Liu2021VoiceFixer} to the setup of this study. We utilize the official implementation\footnote{\url{https://github.com/haoheliu/voicefixer}}. Notably, instead of the HiFi-GAN vocoder employed in the original paper, we use the Vocos vocoder across all systems to ensure consistency. Moreover, as discussed in Section~\ref{sec:related}, several refinement methods have been proposed for specific speech front-ends. We include them as additional front-end-specific baselines. The correspondence between algorithms, training data, and tasks is summarized in Table~\ref{tab:frontend}. Notably, we ensure that all baseline models, including VoiceFixer were trained and evaluated under the identical data conditions as SpeechRefiner for fair comparison.

\begin{table*}[!ht]
\centering

\caption{Performance of SpeechRefiner vs. Task-Specific Refiners Across Front-End Algorithms.}

\begin{tabular}{l|l|c|c|c|c|c|c|c}
\toprule
\textbf{Front-end algorithm} & \textbf{System} & \textbf{COL} & \textbf{DISC} & \textbf{LOUD} & \textbf{NOISE} & \textbf{REVERB} & \textbf{SIG} & \textbf{OVRL} \\ \midrule

\multirow{4}{*}{Enhancement: CDiffuSE} 
& - & 3.16 & 3.86 & 3.56 & 3.44 & 4.10 & 3.20 & 2.85 \\ 
& VoiceFixer & \textbf{3.49} & 4.26 & \textbf{3.94} & 3.89 & 4.45 & 3.57 & 3.23 \\ 
& Diffiner+ & 3.10 & 4.10 & 3.62 & \textbf{4.14} & 4.44 & 3.35 & 3.02 \\ 
\cline{2-9}
& \textbf{SpeechRefiner} & 3.44 & \textbf{4.34} & 3.74 & 4.09 & \textbf{4.55} & \textbf{3.66} & \textbf{3.32} \\ 

\midrule

\multirow{4}{*}{Dereverberation: SGMSE+} 
& - & 3.67 & 3.91 & 4.00 & 3.63 & 4.39 & 3.68 & 3.18 \\ 
& VoiceFixer & 3.78 & 4.03 & 3.95 & 3.85 & 4.52 & 3.46& 3.06 \\ 
& Diffiner+ & 2.79 & 2.83 & 3.10 & 3.19 & 3.49 & 2.43 & 2.20 \\ 
&StoRM & 3.78 & 3.88 & 3.80 & 3.61 & 4.48 & \textbf{3.69} & \textbf{3.23} \\  
\cline{2-9}
& \textbf{SpeechRefiner} & \textbf{3.82} & \textbf{4.17} & \textbf{4.08} & \textbf{3.92} & \textbf{4.60} & 3.60 & 3.20 \\

\midrule

\multirow{4}{*}{Separation: Sepformer} 
&-  & 3.22 & 3.67 & 3.23 & 3.67 & 3.44 & 3.01 & 2.56 \\ 
& VoiceFixer & 3.17 & 3.92 & 3.87 & 3.57 & 4.29 & 3.27 & 2.86 \\
& Fast-Geco & 3.41 & 4.02 & 3.39 & \textbf{3.98} & 3.81 & 3.61 & 3.03 \\
\cline{2-9}
& \textbf{SpeechRefiner}& \textbf{3.89} & \textbf{4.37} & \textbf{4.15} & 3.97 & \textbf{4.56} & \textbf{3.79} & \textbf{3.39} \\
\bottomrule
\end{tabular}
\label{tab:task_specific}
\end{table*}

\subsection{Implementation Details}
\label{ssec:subhead}
All audio files in our datasets are resampled to 24 kHz. The Short-Time Fourier Transform (STFT) is computed using a Hanning window with a window length of 1024 and a hop size of 256. The number of Mel frequency bins is set to 128.

For the Conformer model, we use the AdamW optimizer with \( \beta_1 = 0.9 \), \( \beta_2 = 0.95 \), a weight decay of 0.01, and an initial learning rate of \( 1 \times 10^{-4} \). The model was trained for 500k steps, with a batch-size of $24$.

Vocos is trained for up to 2500k steps with a batch size of 16. The AdamW optimizer is used with a learning rate of  \( 2 \times 10^{-4} \), \( \beta_1 = 0.9 \) and \( \beta_2 = 0.99 \). A cosine annealing scheduler is applied to adjust the learning rate. And the Vocos is trained on entire LibriTTS-R dataset~\cite{Koizumi2023}.

During inference, following~\cite{Popov2021}, a first-order Euler forward ODE solver is used for the flow matching model, and the number of inference steps is set to $64$.

\subsection{Evaluation metrics}
We use SIGMOS\cite{Ristea2024} as our evaluation metric.The SIGMOS is released in the ICASSP 2024 SSI Challenge, which is adhered to the ITU-T P.804 standard. It serves as an objective metric approximating subjective human ratings and assesses speech quality across seven dimensions: Signal (SIG), Coloration (COL), Discontinuity (DISC), Loudness (LOUD), Noisiness (NOISE), Reverb (REVERB), and Overall Quality (OVRL). Trained on over 20,000 speech files, SIGMOS offers a standardized and automated approach for researchers to evaluate speech quality, identify degradation causes, and support studies in speech quality and communication networks. 

\section{Results and Analysis}
\subsection{Comparison with front-end-specific baselines}
We first evaluate SpeechRefiner against state-of-the-art refiners designed for specific front-end tasks. Table~\ref{tab:task_specific} compares the performance across three core tasks: speech enhancement, dereverberation, and separation. Our SpeechRefiner outperforms or matches existing methods in most metrics. Notably, when we tested Diffiner+ (originally designed for speech enhancement) on the dereverberation task using a dereverberation dataset, its performance was significantly poor, with an overall score of 2.20, which highlights the limitation of these models being specifically designed for certain tasks. In contrast, SpeechRefiner achieves consistent improvements, highlighting its broader applicability. For example, on speech separation, SpeechRefiner  achieves an OVRL score of 3.39, surpassing Fast-GeCo (3.03) by 11.9\%. This underscores the advantage of SpeechRefiner’s task-agnostic design, which allows it to generalize across diverse front-end processing tasks without requiring task-specific modifications.

\subsection{Training on the internal dataset}
To further validate the effectiveness of the proposed SpeechRefiner, we utilize our internal front-end processing pipeline to simulate a scenario in which multiple algorithms are applied. In this experiment, we add noise\footnote{Only non-speech noises were selected.} and reverberation from the DNS-Challenge dataset\footnote{\url{https://github.com/microsoft/DNS-Challenge}} to the LibriTTS-R dataset~\cite{Koizumi2023} (Training set merges the $100$, $360$ and $500$ subset, and the test-clean set is used for testing). The distorted data is then processed using the internal front-end pipeline, which consists of denoising, acoustic echo cancellation (AEC), and dereverberation modules. 

\begin{table}[!ht]
\centering
\caption{SpeechRefiner Training on the Internal Dataset and Evaluation Across Front-End Algorithms}
\begin{tabular}{c|c|c|c}
\toprule
\textbf{Front-end algorithm} & \textbf{Refinement} & \textbf{SIG} & \textbf{OVRL} \\ \midrule
\multirow{2}{*}{\makecell[ct]{Enhancement \& Dereverb: \\ Internal}} 
& $\times$ & 3.13 & 2.69 \\
& $\checkmark$ & \textbf{3.72} & \textbf{3.27}  \\ \midrule
\multirow{2}{*}{\makecell[ct]{Enhancement: \\ CDiffuSE}} 
& $\times$ & 3.20 & 2.85 \\ 
& $\checkmark$ & \textbf{3.53} & \textbf{3.13}  \\ \midrule
\multirow{2}{*}{\makecell[ct]{Dereverberation: \\  SGMSE+}} 
& $\times$ &  3.68 & 3.18\\ 
& $\checkmark$ & \textbf{4.06} & \textbf{3.73}\\ \midrule
\multirow{2}{*}{\makecell[ct]{Separation:  \\  Sepformer}} 
& $\times$ &   3.01 & 2.56\\ 
& $\checkmark$&\textbf{ 3.45} & \textbf{3.00}\\ \midrule
\multirow{2}{*}{\makecell[ct]{Audio-TSE:  \\  Spex+ }} 
& $\times$ &  3.04 & 2.64\\ 
& $\checkmark$&\textbf{3.75} & \textbf{3.35}\\ \midrule
\multirow{2}{*}{\makecell[ct]{AV-TSE:  \\ MuSE }} 
& $\times$ &  2.71 & 2.24\\ 
& $\checkmark$&\textbf{3.64} & \textbf{3.19}\\ 
\bottomrule
\end{tabular}
\label{tab:general}
\end{table}

As shown in Table~\ref{tab:general}, SpeechRefiner consistently improves SIG and OVRL scores over unprocessed outputs.  
First, for the test set simulated using the same method as the training data, despite containing distortions introduced by two different front-end processing algorithms, a significant performance improvement is observed. The SIG and OVRL scores increase from 3.13 and 2.69 to 3.72 and 3.27, respectively.  

Additionally, we evaluate SpeechRefiner, trained on the internal dataset, on several unseen test sets from the previous experiment. Compared to unprocessed results, our system consistently demonstrates significant improvements. In some cases, its performance even surpasses that of systems trained on matched training data, highlighting the strong generalization capability of our approach.  

Furthermore, we evaluate the AV-TSE and Audio-TSE and treat them as unseen front-end processing algorithms, including Audio-TSE with Spex+ and AV-TSE with MuSE, where it achieved remarkable performance.

\subsection{Visualization of the refinement effectiveness}
\begin{figure}[!htb]
\centering
\includegraphics[width=0.95\linewidth, height=2\textheight, keepaspectratio]{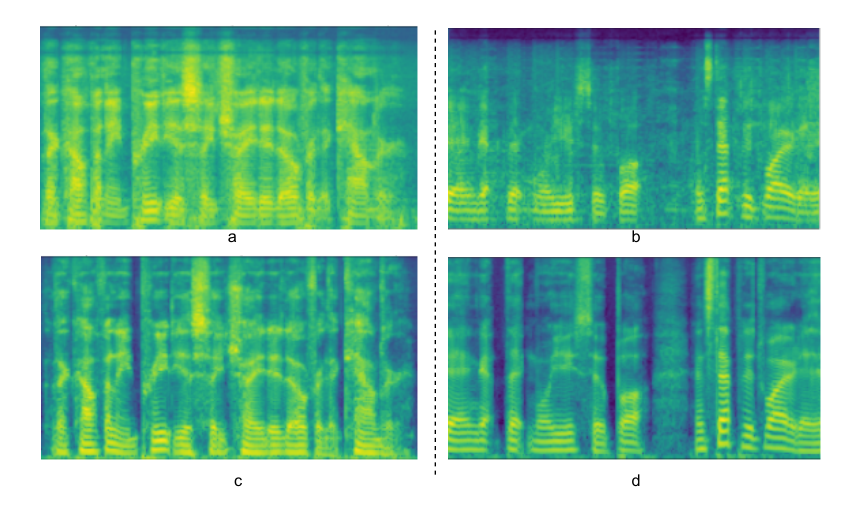}
\caption{Spectrogram Visualization of SpeechRefiner
a) and b) show the outputs of the SGMSE+ and AV-TSE systems, respectively; c) and d) depict the corresponding restored versions produced by SpeechRefiner.}
\label{fig:spec}
\end{figure}

Finally, to better illustrate the effectiveness of the proposed SpeechRefiner, we present the audio spectrograms before and after restoration in Figure~\ref{fig:spec}. By comparing the spectrograms dereverberated by SGMSE+ and extracted by the AV-TSE system with those restored by SpeechRefiner, we observe a significant improvement in clarity. SpeechRefiner effectively eliminates residual reverberation in SGMSE+ outputs and suppresses overlapping speech artifacts in AV-TSE results, confirming its ability to enhance perceptual clarity across diverse scenarios.

\section{Conclusion}
In this paper, we proposed SpeechRefiner, a post-processing module aimed at mitigating impairments and artifacts introduced by various front-end processing algorithms, which significantly affect perceptual quality. Trained on data derived from a single impairment source, SpeechRefiner demonstrated remarkable generalization capabilities across a variety of unseen front-end algorithms.  This underscores the model's robustness despite limited training diversity. Its effectiveness highlights its potential as a complementary tool for improving perceptual quality.

Despite its promising performance, our experiments revealed two main limitations. First, the model struggles with restoring speech containing multiple speakers, often merging them into a single voice. Second, when processing severely distorted speech with limited intelligibility, SpeechRefiner tends to produce clean yet semantically meaningless audio signals.

To address these challenges in future work, we plan to (i) enhance the model's ability to differentiate and restore multi-speaker inputs, (ii) refine its handling of severely degraded speech to preserve semantic accuracy, and (iii) increase the data diversity in the data simulation pipeline to create a more generalizable version of SpeechRefiner for open-sourcing.

\section{Acknowledgements}

This work was supported by National Natural Science Foundation of China, Grant No. 62401377, Shenzhen Science and Technology  Program (Shenzhen Key Laboratory, Grant No. ZDSYS20230626091302006), Shenzhen Science and Technology Research Fund (Fundamental Research Key Project, Grant No. JCYJ20220818103001002),
Program for Guangdong Introducing Innovative and Entrepreneurial Teams, Grant No. 2023ZT10X044.
\bibliographystyle{IEEEtran}
\bibliography{mybib}

\end{document}